\documentclass[a4paper,11pt]{article}
\usepackage{pos}

\def\beqn{\begin{eqnarray}} \def\eeqn{\end{eqnarray}}
\def\beq{\begin{equation}} \def\eeq{\end{equation}}

\title{
\vspace*{-1.5cm}
\begin{minipage}{\textwidth}
{\normalfont\small 
\hspace{\fill} September 2023
}\\
\end{minipage}\\[60pt]
  Tackling Feynman integrals with quantum minimization algorithms}

\ShortTitle{Tackling Feynman integrals with quantum minimization algorithms}

\author*[a,b]{German F. R. Sborlini}

\affiliation[a]{Departamento de F\'isica Fundamental e IUFFyM, Universidad de Salamanca, 37008 Salamanca, Spain.}
\affiliation[b]{Escuela de Ciencias, Ingenier\'ia y Diseño, Universidad Europea de Valencia, Paseo de la Alameda 7, 46010 Valencia, Spain.}

\emailAdd{german.sborlini@usal.es}

\abstract{One of the most severe bottlenecks to reach high-precision predictions in QFT is the calculation of multiloop multileg Feynman integrals. Several new strategies have been proposed in the last years, allowing impressive results with deep implications in particle physics. Still, the efficiency of such techniques starts to drastically decrease when including many loops and legs. In this talk, we explore the implementation of quantum algorithms to optimize the integrands of scattering amplitudes. We rely on the manifestly causal loop-tree duality, which translates the loop into phase-space integrals and avoids the spurious singularities due to non-causal effects. Then, we built a Hamiltonian codifying causal-compatible contributions and minimize it using a Variational Quantum Eigensolver. Our very promising results point towards a potential speed-up for achieving a more numerically-stable representation of Feynman integrals by using quantum computers.}

\FullConference{The European Physical Society Conference on High Energy Physics (EPS-HEP2023)\\
 21-25 August 2023\\
Hamburg, Germany\\}

\begin{document}
\maketitle

\section{Motivation}
\label{sec:Motivation}
In the context of high-energy physics at colliders, we are usually interested in computing cross-sections. Since the energy of the collisions is very high compared to the mass of the particles, we can safely rely on perturbative QFT. Then, the precision of the prediction is increased by adding higher orders to the perturbative expansion. However, this is not an easy task, because we need to deal with Feynman loop integrals and real emission processes. Both contributions contain singularities which must be regulated in order to cancel them. This increases their complexity, leading to severe bottlenecks to go beyond the current state-of-the-art, namely next-to-next-to-leading order (NNLO) in QCD for most of the processes of interest at colliders.

In recent years, several methods were developed to extract very precise predictions from the Standard Model (SM) and other quantum field theories (QFT) for collider phenomenology \cite{Heinrich:2020ybq}. Most of these methods rely on a cancellation of singularities \emph{after} integration although local techniques (i.e. for cancelling the singularities \emph{before} integration) are starting to appear. One of them is based on the Loop-Tree Duality (LTD) theorem \cite{Catani:2008xa,Bierenbaum:2010cy,Bierenbaum:2012th,Buchta:2014dfa}. It allows to open the loops and recast the virtual contributions in terms of tree-level-like amplitudes integrated over a real radiation phase-space. When adding a mapping, both real and virtual contributions can be merged within a single \emph{finite} integral: this is the Four-Dimensional Unsubtraction (FDU) framework \cite{Hernandez-Pinto:2015ysa,Sborlini:2016gbr,Sborlini:2016hat}.

Still, the LTD has shown fascinating properties that go beyond the local cancellation of singularities. In particular, when adding all the dual contributions, the result is manifestly causal \cite{Aguilera-Verdugo:2020set,Aguilera-Verdugo:2020kzc,Ramirez-Uribe:2020hes,Aguilera-Verdugo:2020nrp,Aguilera-Verdugo:2021nrn}. This causal behaviour can be recovered from geometrical rules \cite{Sborlini:2021owe}, and is deeply connected to the identification of directed acyclic graphs (DAG) \cite{Ramirez-Uribe:2021ubp,Clemente:2022nll}.

Besides the development of new methods to calculate cross-sections, overcoming the precision frontier might require to explore novel technologies. One candidate is quantum computing, since it has shown a potential speed-up in database searching \cite{Grover:1997ch,Grover:1997fa}, factorization \cite{Shor:1994jg} and solving minimization/optimization problems \cite{Tilly:2021jem}. In the context of high-energy physics, it has been applied to scattering amplitude calculations \cite{Agliardi:2022ghn,Chawdhry:2023jks,deLejarza:2023qxk}, jet clustering \cite{deLejarza:2022bwc} and, very recently, to perform the causal reconstruction of scattering amplitudes within LTD \cite{Ramirez-Uribe:2021ubp,Clemente:2022nll}. In this article, we will briefly describe how to solve DAG detection with a quantum computer and exploit this information to bootstrap the causal LTD representation of multiloop multileg scattering amplitudes.

\section{Causal Loop-Tree Duality from Geometry}
\label{sec:CausalLTD}
The LTD theorem relies on the iterative application of Cauchy's residue (CR) theorem to remove one degree-of-freedom per loop. Even if the formalism is general, we eliminate the energy component of each loop in order to change the integration domain from a Minkowski to an Euclidean one, thus closely resembling a phase-space integral. In this way, computing the residue is equivalent to put on-shell internal lines of the corresponding loop diagram.

The iterative application of the CR theorem has several subtleties \cite{Aguilera-Verdugo:2020nrp}, among them the cancellation of non-physical contributions originated from the so-called \emph{displaced poles}. In the end, each combination of poles contributing to the dual amplitude (i.e. the LTD representation of the amplitude) can be mapped into a tree-level-like diagram obtained by cutting $L$ lines for a $L$-loop Feynman amplitude.

Furthermore, when all the dual contributions are added together, a noticeable simplification takes place \cite{Aguilera-Verdugo:2020set,Aguilera-Verdugo:2020kzc,Aguilera-Verdugo:2020nrp}. In fact, any $L$-loop scattering amplitude with $V$ vertices can be expressed as
\beq 
A^{(L)} = \sum_{\sigma \in \Sigma} \int_{\vec{\ell}_1, \ldots, \vec{\ell}_1} (-1)^k\, \frac{{\cal N}_{\sigma}(\{q_{r,0}^{(+)}\},\{p_{j,0}\})}{x_{L+k}} \, \prod_{i=1}^k \frac{1}{\lambda_{\sigma(i)}} \, + \, (\sigma \longleftrightarrow \bar{\sigma}) \, ,
\label{eq:Causal}
\eeq
which is known as a \emph{causal representation}. In this expression, $k=V-1$ is the order of the topology associated to the \emph{reduced} Feynman diagram, $q^{(+)}_{i,0}$ is the on-shell positive energy of the $i$-th internal line and $x_{L+k} = \prod_{j=1}^{L+k} 2 q^{(+)}_{j,0}$ is the analogous to the integration measure of the phase-space\footnote{More information about this formula is available in Refs. \cite{Aguilera-Verdugo:2020set,TorresBobadilla:2021ivx,Sborlini:2021owe}.}. $\{\lambda_i\}$ are associated to physical threshold singularities of the diagram and $1/\lambda_i$ are known as \emph{causal propagators}. $\sigma$ represents a combination of $k$ \emph{entangled thresholds}, and the compatibility conditions that determine $\Sigma$ can be fixed by following geometrical rules \cite{Sborlini:2021owe}. One of these rules establishes that the binary partitions associated to each $\lambda_{\sigma(i)}$ for $\sigma \in \Sigma$ must have a compatible momentum flux. It turns out that such momentum flux compatibility is possible if and only if the associated directed graph (DG) for the reduced Feynman diagram has no loops \cite{Ramirez-Uribe:2021ubp,Sborlini:2021cba}. Thus, the geometrical formulation of causal LTD implies that the causal representation in Eq. (\ref{eq:Causal}) can be bootstrapped from DAGs.

\section{Quantum algorithms for LTD}
\label{sec:QuantumAlgorithms}
At this point, the problem of finding a causal representation has been partially reduced to the identification of DAGs. Classical algorithms for this purpose rely on construction strategies, as well as the direct evaluation of acyclicity conditions. In both cases, for dense graphs (i.e. corresponding to multiloop multileg Feynman diagrams with multi-particle interactions), the scaling is almost exponential in the number of edges. Being able to efficiently parallelize the evaluation of the acyclicity condition could lead to an important speed-up: for this reason, we decided to explore quantum algorithms.

\begin{figure}[ht!]
    \centering
    \includegraphics[scale=0.44]{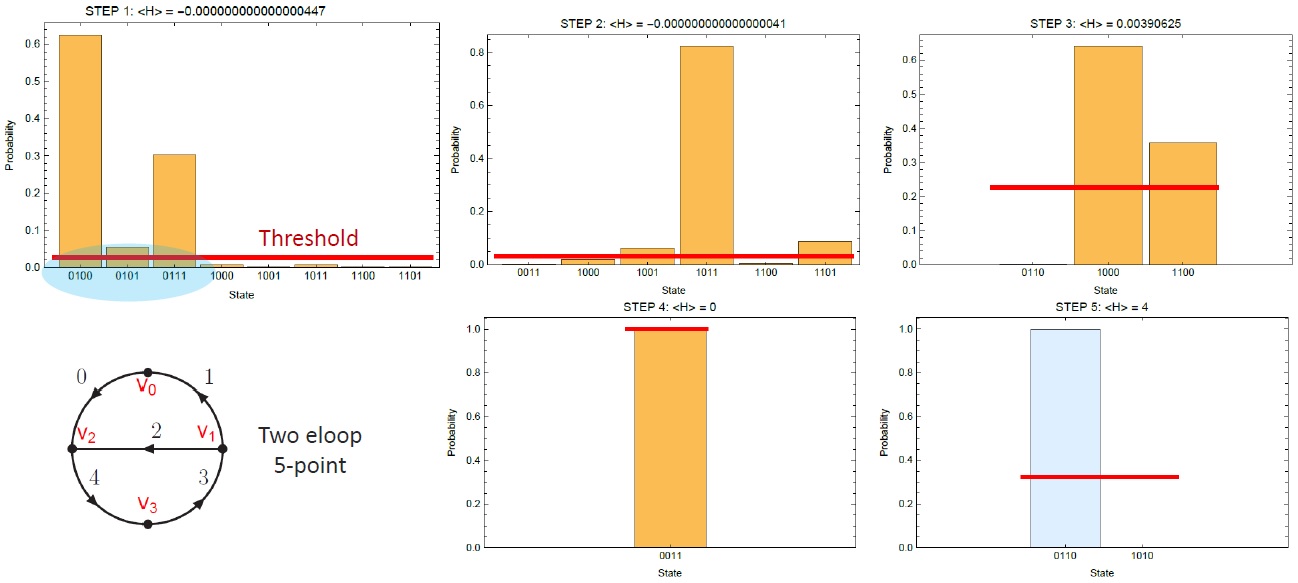}
    \caption{Execution of the multi-run VQE algorithm on a representative topology with 5 edges. In four steps, we manage to collect exactly the 9 acyclic configurations, with 0 error rate.}
    \label{fig:EJEMPLO}
\end{figure} 

Let us start considering a directed graph $G$ composed by the vertices $V$ linked by the set of edges $E$. Each element in $E$ is denoted as $e_i=(v_{i,0},v_{i,1})$. To define a quantum algorithm, we promote the vertices and edges to a Hilbert space ${\cal H}=E \times V$. From graph theory, we know that all the geometric information of $G$ is codified within the adjacency matrix $A$, so we define the associated operator
\beq
{\cal A} = \sum_{e_i=(v_{i,0},v_{i,1}) \in E} \left[\sigma_{v_0}^- \pi^0_e \sigma_{v_1}^+ + \sigma_{v_1}^- \pi^1_e \sigma_{v_0}^+ \right] \, ,
\label{eq:Aoperator}
\eeq
with $\sigma^\pm_v$ the Pauli matrices in the vertex space and $\pi_e^{0,1}$ a projector w.r.t. a reference orientation of the edges \cite{Clemente:2022nll}. The powers of the classical adjacency matrix contain information about cycles: if ${\rm tr}(A^n)=0$ then there are not $n$-cycles. So, we define the Hamiltonian
\beq
{\cal H} = \sum_{n=1}^{\#(V)} {\rm tr}_V\left({\cal A}^n\right) \, ,
\label{eq:Hoperator}
\eeq
that will have, by construction, a degenerated ground state with zero energy composed by all the possible acyclic configurations of $G$. This Hamiltonian is an operator acting on the space of edges, since we trace over the space of vertices. 

In Ref. \cite{Clemente:2022nll}, we built Hamiltonians for different topologies and minimized it using a Variational Quantum Eigensolver (VQE) \cite{Tilly:2021jem}, which can offer advantages w.r.t. classical minimization algorithms. However, standard VQE is not intended to minimize a multi-degenerated problem and, in our case, there are roughly $2^{\#E -1}$ solutions for some topologies. For this reason, we proposed a \emph{multi-run} VQE algorithm: we run a first VQE, impose a threshold to retain potential solutions, add a penalization term to ${\cal H}$ and then re-run the VQE. We use the energy of the approximated ground state to control the converge and penalization terms to avoid finding repeated solutions. In Fig. \ref{fig:EJEMPLO} we present a representative example in which our algorithm took 4 steps to collect all the acyclic graphs associated to a sunrise-like diagram.

\section{Conclusions and outlook}
\label{sec:conclusions}
In this article, we review potential advantages of the LTD theorem and the corresponding causal representations for computing multiloop multileg scattering amplitudes. The causal representation contains only physical singularities at integrand level, and it is defined in an Euclidean space. As we shown in Ref. \cite{Sborlini:2021owe}, it can be obtained through geometrical selection rules that provide the full set of causal-compatible entangled thresholds. Furthermore, in Refs. \cite{Ramirez-Uribe:2021ubp,Clemente:2022nll}, we show that causal representations can be bootstrapped from the set of DAGs associated to the Feynman diagrams.

It turns out that relating causality and DAGs allows to recast the problem of obtaining causal representations in such a way that it can be handled by quantum computers. Thus, we codified the possible directions of the edges into qubits and we define a Hamiltonian whose ground state contains all the possible acyclic graphs. By using a multi-run VQE algorithm, we manage to identify the acyclic graphs with a high success rate. These developments can be considered as a first step towards computing Feynman integrals through quantum minimization algorithms. Besides allowing to calculate complicated multiloop multileg scattering amplitudes, these quantum algorithms might revolutionize the way in which we obtain high-precision predictions from QFT.

  \subsection*{Acknowledgments}
We thank D. Suarez Fontanella, D. Renter\'ia-Estrada and A. Renter\'ia-Olivo for their fruitful comments. This work is supported by the Spanish Government (Agencia Estatal de Investigaci\'on MCIN /AEI/10.13039/501100011033) Grants No. PID2020-114473GB-I00, PID2022-141910NB-I00, Generalitat Valenciana Grant No. PROMETEO/2021/071 and H2020-MSCA-COFUND USAL4EXCELLENCE-PROOPI-391 project under Grant Agreement No 101034371.


\begin{thebibliography}{10}

\bibitem{Heinrich:2020ybq}
G.~Heinrich, \emph{{Collider Physics at the Precision Frontier}},
  \href{http://dx.doi.org/10.1016/j.physrep.2021.03.006}{\emph{Phys. Rept.}
  {\bf 922} (2021) 1--69}, [\href{http://arxiv.org/abs/2009.00516}{{\tt
  2009.00516}}].

\bibitem{Catani:2008xa}
S.~Catani, T.~Gleisberg, F.~Krauss, G.~Rodrigo and J.-C. Winter, \emph{{From
  loops to trees by-passing Feynman's theorem}},
  \href{http://dx.doi.org/10.1088/1126-6708/2008/09/065}{\emph{JHEP} {\bf 09}
  (2008) 065}, [\href{http://arxiv.org/abs/0804.3170}{{\tt 0804.3170}}].

\bibitem{Bierenbaum:2010cy}
I.~Bierenbaum, S.~Catani, P.~Draggiotis and G.~Rodrigo, \emph{{A Tree-Loop
  Duality Relation at Two Loops and Beyond}},
  \href{http://dx.doi.org/10.1007/JHEP10(2010)073}{\emph{JHEP} {\bf 10} (2010)
  073}, [\href{http://arxiv.org/abs/1007.0194}{{\tt 1007.0194}}].

\bibitem{Bierenbaum:2012th}
I.~Bierenbaum, S.~Buchta, P.~Draggiotis, I.~Malamos and G.~Rodrigo,
  \emph{{Tree-Loop Duality Relation beyond simple poles}},
  \href{http://dx.doi.org/10.1007/JHEP03(2013)025}{\emph{JHEP} {\bf 03} (2013)
  025}, [\href{http://arxiv.org/abs/1211.5048}{{\tt 1211.5048}}].

\bibitem{Buchta:2014dfa}
S.~Buchta, G.~Chachamis, P.~Draggiotis, I.~Malamos and G.~Rodrigo, \emph{{On
  the singular behaviour of scattering amplitudes in quantum field theory}},
  \href{http://dx.doi.org/10.1007/JHEP11(2014)014}{\emph{JHEP} {\bf 11} (2014)
  014}, [\href{http://arxiv.org/abs/1405.7850}{{\tt 1405.7850}}].

\bibitem{Hernandez-Pinto:2015ysa}
R.~J. Hernandez-Pinto, G.~F.~R. Sborlini and G.~Rodrigo, \emph{{Towards gauge
  theories in four dimensions}},
  \href{http://dx.doi.org/10.1007/JHEP02(2016)044}{\emph{JHEP} {\bf 02} (2016)
  044}, [\href{http://arxiv.org/abs/1506.04617}{{\tt 1506.04617}}].

\bibitem{Sborlini:2016gbr}
G.~F.~R. Sborlini, F.~Driencourt-Mangin, R.~Hernandez-Pinto and G.~Rodrigo,
  \emph{{Four-dimensional unsubtraction from the loop-tree duality}},
  \href{http://dx.doi.org/10.1007/JHEP08(2016)160}{\emph{JHEP} {\bf 08} (2016)
  160}, [\href{http://arxiv.org/abs/1604.06699}{{\tt 1604.06699}}].

\bibitem{Sborlini:2016hat}
G.~F.~R. Sborlini, F.~Driencourt-Mangin and G.~Rodrigo, \emph{{Four-dimensional
  unsubtraction with massive particles}},
  \href{http://dx.doi.org/10.1007/JHEP10(2016)162}{\emph{JHEP} {\bf 10} (2016)
  162}, [\href{http://arxiv.org/abs/1608.01584}{{\tt 1608.01584}}].

\bibitem{Aguilera-Verdugo:2020set}
J.~J. Aguilera-Verdugo, F.~Driencourt-Mangin, R.~J. Hern\'andez-Pinto,
  J.~Plenter, S.~Ramirez-Uribe, A.~E. Renteria~Olivo et~al., \emph{{Open Loop
  Amplitudes and Causality to All Orders and Powers from the Loop-Tree
  Duality}},
  \href{http://dx.doi.org/10.1103/PhysRevLett.124.211602}{\emph{Phys. Rev.
  Lett.} {\bf 124} (2020) 211602}, [\href{http://arxiv.org/abs/2001.03564}{{\tt
  2001.03564}}].

\bibitem{Aguilera-Verdugo:2020kzc}
J.~J. Aguilera-Verdugo, R.~J. Hernandez-Pinto, G.~Rodrigo, G.~F.~R. Sborlini
  and W.~J. Torres~Bobadilla, \emph{{Causal representation of multi-loop
  Feynman integrands within the loop-tree duality}},
  \href{http://dx.doi.org/10.1007/JHEP01(2021)069}{\emph{JHEP} {\bf 01} (2021)
  069}, [\href{http://arxiv.org/abs/2006.11217}{{\tt 2006.11217}}].

\bibitem{Ramirez-Uribe:2020hes}
S.~Ram\'\i{}rez-Uribe, R.~J. Hern\'andez-Pinto, G.~Rodrigo, G.~F.~R. Sborlini
  and W.~J. Torres~Bobadilla, \emph{{Universal opening of four-loop scattering
  amplitudes to trees}},
  \href{http://dx.doi.org/10.1007/JHEP04(2021)129}{\emph{JHEP} {\bf 04} (2021)
  129}, [\href{http://arxiv.org/abs/2006.13818}{{\tt 2006.13818}}].

\bibitem{Aguilera-Verdugo:2020nrp}
J.~Aguilera-Verdugo, R.~J. Hern\'andez-Pinto, G.~Rodrigo, G.~F.~R. Sborlini and
  W.~J. Torres~Bobadilla, \emph{{Mathematical properties of nested residues and
  their application to multi-loop scattering amplitudes}},
  \href{http://dx.doi.org/10.1007/JHEP02(2021)112}{\emph{JHEP} {\bf 02} (2021)
  112}, [\href{http://arxiv.org/abs/2010.12971}{{\tt 2010.12971}}].

\bibitem{Aguilera-Verdugo:2021nrn}
J.~Aguilera-Verdugo et~al., \emph{{A Stroll through the Loop-Tree Duality}},
  \href{http://dx.doi.org/10.3390/sym13061029}{\emph{Symmetry} {\bf 13} (2021)
  1029}, [\href{http://arxiv.org/abs/2104.14621}{{\tt 2104.14621}}].

\bibitem{Sborlini:2021owe}
G.~F.~R. Sborlini, \emph{{Geometrical approach to causality in multiloop
  amplitudes}},
  \href{http://dx.doi.org/10.1103/PhysRevD.104.036014}{\emph{Phys. Rev. D} {\bf
  104} (2021) 036014}, [\href{http://arxiv.org/abs/2102.05062}{{\tt
  2102.05062}}].

\bibitem{Ramirez-Uribe:2021ubp}
S.~Ram\'\i{}rez-Uribe, A.~E. Renter\'\i{}a-Olivo, G.~Rodrigo, G.~F.~R. Sborlini
  and L.~Vale~Silva, \emph{{Quantum algorithm for Feynman loop integrals}},
  \href{http://dx.doi.org/10.1007/JHEP05(2022)100}{\emph{JHEP} {\bf 05} (2022)
  100}, [\href{http://arxiv.org/abs/2105.08703}{{\tt 2105.08703}}].

\bibitem{Clemente:2022nll}
G.~Clemente, A.~Crippa, K.~Jansen, S.~Ram\'\i{}rez-Uribe, A.~E.
  Renter\'\i{}a-Olivo, G.~Rodrigo et~al., \emph{{Variational quantum
  eigensolver for causal loop Feynman diagrams and acyclic directed graphs}},
  \href{http://arxiv.org/abs/2210.13240}{{\tt 2210.13240}}.

\bibitem{Grover:1997ch}
L.~K. Grover, \emph{{Quantum computers can search rapidly by using almost any
  transformation}},
  \href{http://dx.doi.org/10.1103/PhysRevLett.80.4329}{\emph{Phys. Rev. Lett.}
  {\bf 80} (1998) 4329--4332},
  [\href{http://arxiv.org/abs/quant-ph/9712011}{{\tt quant-ph/9712011}}].

\bibitem{Grover:1997fa}
L.~K. Grover, \emph{{Quantum mechanics helps in searching for a needle in a
  haystack}}, \href{http://dx.doi.org/10.1103/PhysRevLett.79.325}{\emph{Phys.
  Rev. Lett.} {\bf 79} (1997) 325--328},
  [\href{http://arxiv.org/abs/quant-ph/9706033}{{\tt quant-ph/9706033}}].

\bibitem{Shor:1994jg}
P.~W. Shor, \emph{{Polynomial time algorithms for prime factorization and
  discrete logarithms on a quantum computer}},
  \href{http://dx.doi.org/10.1137/S0097539795293172}{\emph{SIAM J. Sci.
  Statist. Comput.} {\bf 26} (1997) 1484},
  [\href{http://arxiv.org/abs/quant-ph/9508027}{{\tt quant-ph/9508027}}].

\bibitem{Tilly:2021jem}
J.~Tilly et~al., \emph{{The Variational Quantum Eigensolver: A review of
  methods and best practices}},
  \href{http://dx.doi.org/10.1016/j.physrep.2022.08.003}{\emph{Phys. Rept.}
  {\bf 986} (2022) 1--128}, [\href{http://arxiv.org/abs/2111.05176}{{\tt
  2111.05176}}].

\bibitem{Agliardi:2022ghn}
G.~Agliardi, M.~Grossi, M.~Pellen and E.~Prati, \emph{{Quantum integration of
  elementary particle processes}},
  \href{http://dx.doi.org/10.1016/j.physletb.2022.137228}{\emph{Phys. Lett. B}
  {\bf 832} (2022) 137228}, [\href{http://arxiv.org/abs/2201.01547}{{\tt
  2201.01547}}].

\bibitem{Chawdhry:2023jks}
H.~A. Chawdhry and M.~Pellen, \emph{{Quantum simulation of colour in
  perturbative quantum chromodynamics}},
  \href{http://arxiv.org/abs/2303.04818}{{\tt 2303.04818}}.

\bibitem{deLejarza:2023qxk}
J.~J.~M. de~Lejarza, M.~Grossi, L.~Cieri and G.~Rodrigo, \emph{{Quantum Fourier
  Iterative Amplitude Estimation}},
  \href{http://arxiv.org/abs/2305.01686}{{\tt 2305.01686}}.

\bibitem{deLejarza:2022bwc}
J.~J.~M. de~Lejarza, L.~Cieri and G.~Rodrigo, \emph{{Quantum clustering and jet
  reconstruction at the LHC}},
  \href{http://dx.doi.org/10.1103/PhysRevD.106.036021}{\emph{Phys. Rev. D} {\bf
  106} (2022) 036021}, [\href{http://arxiv.org/abs/2204.06496}{{\tt
  2204.06496}}].

\bibitem{TorresBobadilla:2021ivx}
W.~J. Torres~Bobadilla, \emph{{Loop-tree duality from vertices and edges}},
  \href{http://dx.doi.org/10.1007/JHEP04(2021)183}{\emph{JHEP} {\bf 04} (2021)
  183}, [\href{http://arxiv.org/abs/2102.05048}{{\tt 2102.05048}}].

\bibitem{Sborlini:2021cba}
G.~F.~R. Sborlini, \emph{{Geometry and causal flux in multi-loop Feynman
  diagrams}},
  \href{http://dx.doi.org/10.31349/SuplRevMexFis.3.020703}{\emph{Rev. Mex. Fis.
  Suppl.} {\bf 3} (2022) 020703}, [\href{http://arxiv.org/abs/2109.14476}{{\tt
  2109.14476}}].

\end{thebibliography}

\providecommand{\href}[2]{#2}\begingroup\raggedright\endgroup

\end{document}